\documentclass[twocolumn,aps,pre,amsmath,showpacs]{revtex4-1}
\usepackage{graphicx,dcolumn,bm,amssymb,amsmath,ulem,indentfirst,amsthm}
\usepackage[toc,page]{appendix}
\usepackage{color}

\theoremstyle{plain}
\def\be{\begin{equation}}
\def\ee{\end{equation}}

\newtheorem*{theorem*}{Theorem}

\begin{document}

\author{Bingyu Cui$^{1,2}$, Rico Milkus$^{1}$, Alessio Zaccone$^{1,3}$}
\affiliation{${}^1$Statistical Physics Group, Department of Chemical
Engineering and Biotechnology, University of Cambridge, New Museums Site, CB2
3RA Cambridge, U.K.}
\affiliation{${}^2$Department of Applied Mathematics and Theoretical Physics,
University of Cambridge,
Wilberforce Road, Cambridge CB3 0WA, U.K.}
\affiliation{${}^3$Cavendish Laboratory, University of Cambridge, JJ Thomson
Avenue, CB30HE Cambridge,
U.K.}
\begin{abstract}

Amorphous solids or glasses are known to exhibit stretched-exponential decay over broad time intervals
in several of their macroscopic observables: intermediate scattering function, dielectric relaxation modulus, time-elastic modulus etc.
This behaviour is prominent especially near the glass transition. In this Letter we show, on the example of dielectric relaxation,
that stretched-exponential relaxation is intimately related to the peculiar lattice dynamics of glasses. By reformulating the
Lorentz model of dielectric matter in a more general form, we express the dielectric response as a function of the vibrational density of states (DOS)
for a random assembly of spherical particles interacting harmonically with their nearest-neighbours. Surprisingly we find that near
the glass transition for this system (which coincides with the Maxwell rigidity transition), the dielectric relaxation is perfectly consistent with
stretched-exponential behaviour with Kohlrausch exponents $0.56 < \beta < 0.65$, which is the range where exponents are measured in most experimental systems.
Crucially, the root cause of stretched-exponential relaxation can be traced back to soft modes (boson-peak) in the DOS.

\end{abstract}

\title{The relation between stretched-exponential relaxation and the vibrational density of states in glassy disordered systems}
\maketitle

\section{Introduction}
Since its first observation by Kohlrausch in 1847~\cite{Cardona}, stretched-exponential relaxation has been observed in the time-dependent relaxation
of several (elastic, dielectric, electronic) macroscopic observables in nearly all structurally disordered solids. Ultimately, this behaviour represents one of the most common hallmarks of irreversibility in disordered systems.
Over the last century, stretched-exponentials have been used in countless experimental settings to fit experimental data. Although it is common knowledge that
stretched-exponential relaxation relates somehow to spatially heterogeneous many-body interactions or to heterogeneous distribution of activation energy barriers
~\cite{Montroll,Langer}, only very few models or theories are able to predict stretched-exponential relaxation from first-principle dynamics~\cite{Phillips}. In fact, strictly speaking, only two models recover stretched-exponential relaxation in well-defined specific situations. One is a model of electronic relaxation via non-radiative exciton-hole recombination where holes are randomly distributed traps that "eat up" the diffusing excitons~\cite{Lifshitz,Luttinger,Procaccia}. As shown in Ref.~\cite{Procaccia}, according to this model the density of not-yet trapped excitons decays at long times as $\sim \exp(-t^{d/d+2})$, which gives a Kohlrausch exponent $\beta=0.6$ in 3d. In spite of the elegance of this model, it is not straightforward to apply it to elucidate Kohlrausch relaxation in glasses.
The other model is the Mode-Coupling Theory (MCT) of supercooled liquids, which gives a solution that can be approximated with a stretched-exponential for the intermediate scattering function at temperatures well above the glass transition~\cite{Goetze, Mazenko, Goetze-book}.

Clearly, these two models are quite specific and limited in their applicability~\cite{Donth,Ngai}. For example, starting with Kohlrausch original experiment using the Leyden glass jar, most of the physical systems where stretched-exponential behaviour has been observed are represented by disordered solids, well below the glass transition where MCT is no longer applicable. This is a very important topic in electrical engineering, where dielectric insulators in the solid state are typically employed for all high-voltage transmission applications~\cite{Kao}.

Hence, while stretched-exponential relaxation is ubiquitous in the solid-state, it has not been possible to trace it back to a well-defined mechanism in the many-body dynamics, or to a well-defined microscopic descriptor of the dynamics.
In this Letter, we re-examine the problem from the point of view of lattice dynamics, suitably evaluated for a model of disordered solids. Focusing on the paradigmatic case of dielectric relaxation, it is shown that stretched-exponential decay of the dielectric modulus over many decades in time is recovered by the numerical solution to the Lorentz sum-rule with a DOS that takes into account the crucial role of so-called boson-peak modes over the Debye $\sim \omega_{p}^{2}$ law. Since the origin of these modes lies in the Ioffe-Regel crossover~\cite{Ioffe,Tanaka,Schirmacher,Parshin} at which phonons are scattered off by the disorder (in particular, by the absence of inversion-symmetry of the lattice~\cite{Milkus}), it is possible to establish a direct link between stretched-exponential relaxation and the quasi-localization of phonons by the disorder at the boson peak.

\section{Goldhaber-Teller model of disordered dielectric}
In the following, we work within the assumption of
disordered elastically bound classical charges, which is the same as the Goldhaber-Teller model originally developed to explain the giant dipole resonance in atomic nuclei~\cite{Teller, Myers}.
In this model, schematically depicted in Fig.1, two types of charges, positive and negative (or, equivalently, positive and neutral, as in the Goldhaber-Teller model for nuclei), are inter-dispersed randomly in space. Every positive charge is surrounded by nearest-neighbours (which are negative, or neutral), to each of which it is bound by an attractive harmonic potential.
In a dielectric solid or in a supercooled ionic liquid~\cite{Richert}, an attraction minimum around which harmonic approximation can be taken, may come from a superposition of electrostatic attraction and van der Waals attraction, competing with short-range steric repulsion (while in the original Goldhaber-Teller model the attraction comes, evidently, from the strong nuclear force).

\begin{figure}
\begin{center}
\includegraphics[height=5.7cm,width=9.2cm]{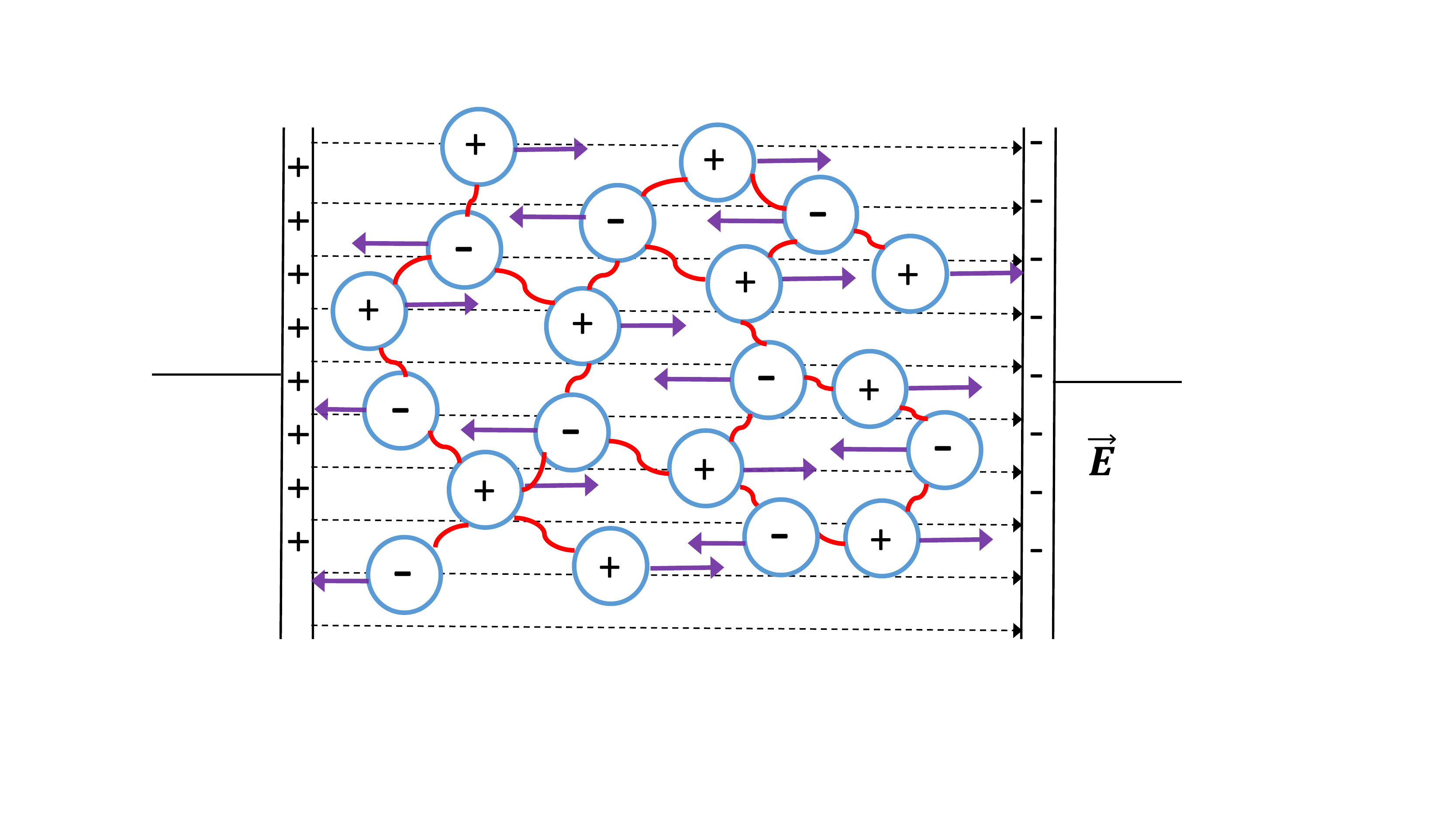}
\caption{Goldhaber-Teller model of a disordered dielectric solid. Spherical particles with positive and negative charge (or equivalently, neutral) are inter-dispersed randomly. Each particle interacts with its nearest-neighbours (which bear opposite charge) uniquely via an attractive harmonic potential (linear springs, in red). Note that the charges do not contribute explicitly to the harmonic inter-particle interaction.}
\end{center}
\end{figure}

In order to evaluate the dielectric response (below) on the basis of lattice dynamics,
we make use of a DOS $\rho(\omega_{p})$ obtained by numerical diagonalization
of a model random lattice of harmonically-bound spherical particles. This random network is obtained by driving a
dense Lennard-Jones liquid into a metastable glassy energy minimum with a
Monte-Carlo relaxation algorithm, and then replacing all the nearest-neighbour
pairs with harmonic springs all of the same spring
constant, and with a relatively narrow spring-length distribution~\cite{Milkus}. Springs are then cut at random in the lattice to
generate random lattices with variable mean coordination $Z$, from
$Z=9$ down to the isostatic Maxwell limit $Z=2d=6$. It is important to notice that this simplified model DOS
is applicable only to systems where the building blocks are spherical and interact with central-force potentials.
The DOS obtained from numerical diagonalization of the simulated network is
expressed in terms of dimensionless eigenfrequencies $\omega_{p}$.


In this random assembly of particles, only nearest-neighbour interactions are present, and the number $Z$ represents the average coordination number or
average number of nearest-neighbours per particle.
Also, the DOS obtained from diagonalization of the model random networks,
depends sensitively on the average coordination number $Z$. For example, the boson peak
frequency drifts towards lower values of $\omega_{p}$ upon increasing $Z$, according to the scaling
$\omega_{p}^{BP}\sim (Z-6)$, as observed also in random packings~\cite{Silbert}. This behaviour is also consistent with the common observation that in glasses the
boson peak frequency shifts to lower frequency upon increasing the density or the pressure~\cite{Alba-Simionesco}. Hence, $Z$ is the crucial control parameter of the
relaxation process, which, in a real e.g. molecular glass, changes with $T$. Therefore,
in order to use our numerical DOS data in the evaluation of the dielectric
function, we need to find a physically meaningful relation between $Z$ and $T$
at the glass transition.

In all experimental systems which measure the $T$-dependent material response,
the temperature is varied at constant pressure, which implies that thermal
expansion is important. Following previous work, we thus employ thermal
expansion ideas~\cite{Zaccone2013} to relate $Z$ and $T$. Upon introducing the
thermal expansion coefficient $\alpha_T=\frac{1}{V}(\partial{V}/\partial{T})$
and replacing the total volume $V$ of the sample via the volume fraction $\phi=v
N/V$ occupied by the molecules ($v$ is the volume of one molecule), upon
integration we obtain $\ln{(1/\phi)}=\alpha_T T + const$. Approximating $Z \sim
\phi$ locally, we get $Z=Z_{0} e^{-\alpha_T T}$. Imposing that  $Z_{0}=12$, as
for FCC crystals at $T=0$ in accordance with Nernst third-law principle, we finally get,
for the example of
glycerol, $Z\approx 6.02$ when $T=184K$. This is very close to the reported
$T_{g}$ for this material~ \cite{Lunkenheimer}. For $Z=7,8$ and $9$, the corresponding
temperatures are calculated to be $T=144~K, 108~K$ and $77~K$.

\begin{figure}
\begin{center}
\includegraphics[height=5.7cm,width=8.7cm]{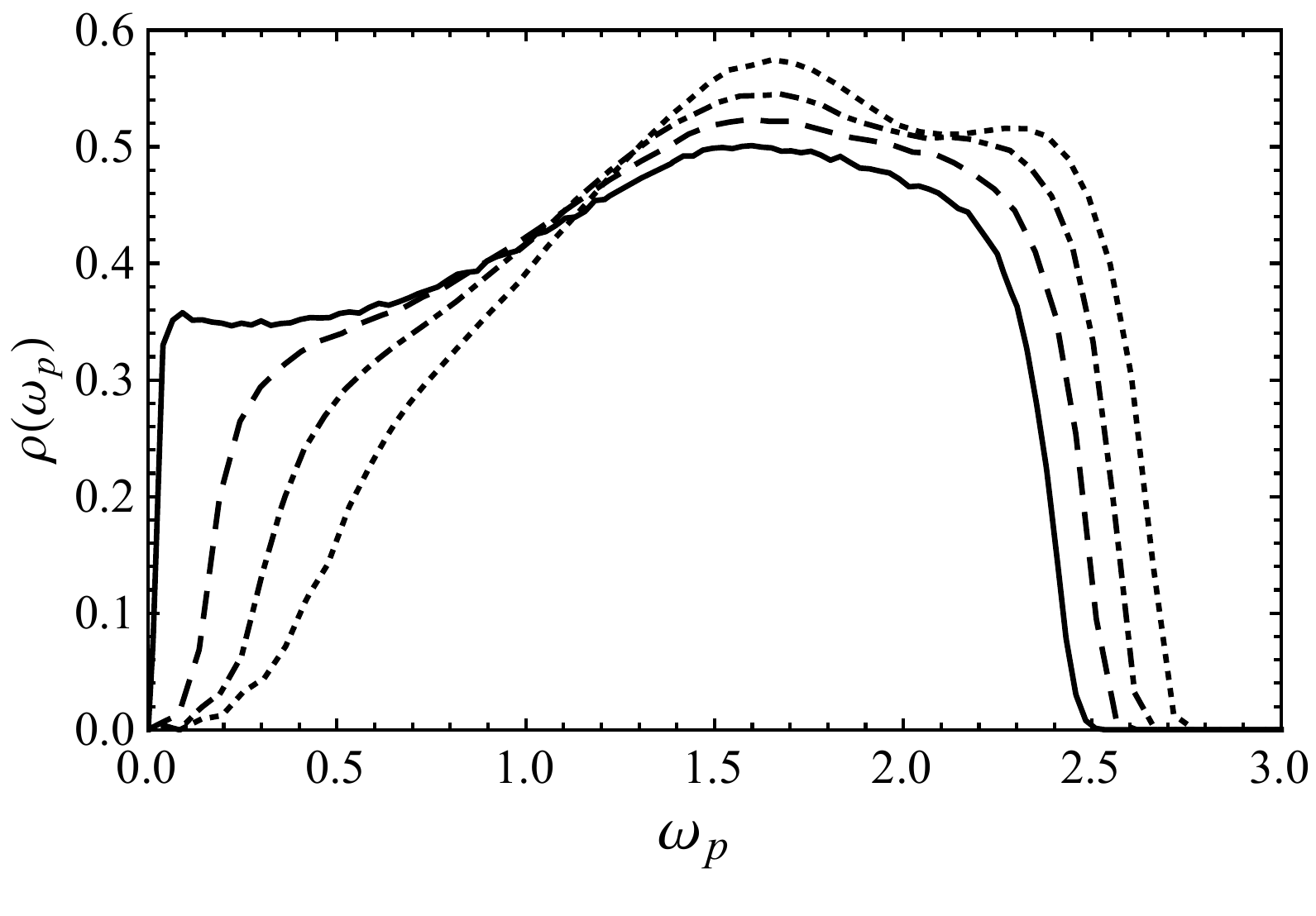}
\caption{Density of states (DOS) with respect to eigenfrequency $\omega_{p}$ at $Z=6.1$, i.e. close to the marginal stability limit $Z=6$ that we identify here as the solid-liquid (glass) transition, $Z=7, Z=8, Z=9$, which are marked as solid, dashed, dotdashed and dotted lines respectively.}
\end{center}
\end{figure}

\begin{figure}
\begin{center}
\includegraphics[height=5.7cm,width=8.7cm]{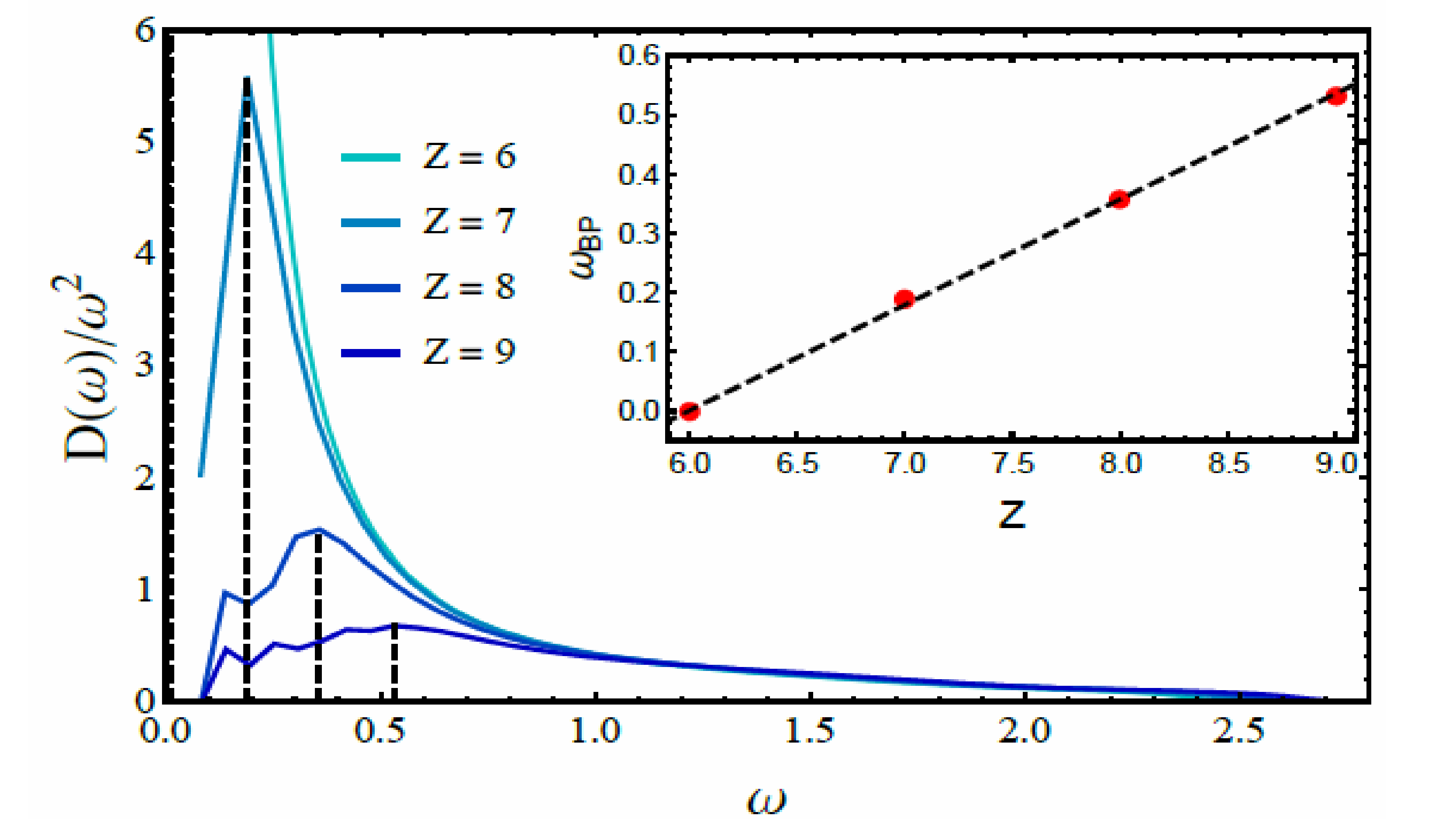}
\caption{The DOS normalized by Debye's $\omega_p^2$ law, for (from bottom to top): $Z=9, Z=8, Z=7, Z=6$, which gives evidence of the boson peak at low $\omega_p$. The eigenfrequency of boson peak scales as $\omega_p^{\textit{BP}}\sim(Z-6)$ as known from work for disordered systems with cental-force interactions~\cite{Silbert,Milkus}.}
\end{center}
\end{figure}

It is seen in Fig.2 that for the case $Z=6.1$, i.e. very close to the solid-liquid rigidity transition that occurs at $Z=6$, a strong boson peak is present
in the DOS. The continuum Debye regime $\sim \omega_{p}^{2}$ is not visible or
absent, whereas an infinitesimal gap between
$\omega_{p}=0$ and the lowest eigenfrequency exists. Hence, under conditions close to the glass transition, the
vibrational spectrum is dominated by soft modes at low frequency. Upon increasing $Z$, the extent of soft modes at low frequency decreases markedly while the Debye $\sim\omega_{p}^{2}$ regime extends to higher frequencies. At the highest $Z$ values, the relics of van Hove singularities that characterize FCC crystals become visible, because medium-range order  has to increase upon increasing the coordination $Z$~\cite{Milkus}, towards the FCC limit $Z=12$. In particular, the local degree of centrosymmetry of the nearest-neighbours is the crucial form of order which increases upon increasing $Z$ and correlates directly with the boson peak~\cite{Milkus}. 
Note that $Z_0=12$ for FCC is independent of density only if the thermal fluctuations are neglected (as is the case at $T=0$ or for athermal systems, e.g. hard spheres). At a finite $T$, defects are created by thermal excitations which may lower the total average $Z$. Upon further increasing $T$, amorphous regions are formed with $Z_0<12$, until the FCC crystal either melts or undergoes a transition into a fully amorphous glass (crystal-glass transition). 
In Fig.3, the increase of the boson peak (normalized by the Debye level $\sim \omega_{p}^{2}$) upon decreasing $Z$ is plotted, and the boson-peak frequency dependence on $Z$ is shown in the inset.

\section{Dielectric response of the Goldhaber-Teller disordered lattice model}
In order to determine the dependence of the polarization and of the dielectric function on the frequency of the field, we have to describe the displacement $\uline{r}$ of each molecule from its own equilibrium position under the applied field $\uline{E}$. Upon treating the dynamics classically, the equation of motion for a charge $i$ under the forces coming from interactions with other charges and from the applied electric filed, is given by the following phenomenological Lorentz damped oscillator equation~\cite{Born-Wolf}
\begin{equation}
m\ddot{ \uline{r}}_{i}+\nu\dot{\uline{r}}_{i}+\uline{\uline{H}}_{ij}\uline{r}_{j}=q\uline{E}
\end{equation}
where $m$ is the (effective) mass of the charged group, and  $q$ is its net electric charge. The Hessian $\uline{\uline{H}}_{ij}=\partial U/\partial \uline{r}_{i}\partial\uline{r}_{j}$, where $U$ is the total potential energy of the system, represents the restoring attractive interactions from oppositely-charged nearest-neighbour charges, that tend to bring the charge $i$ back to the rest position that $i$ had at zero-field. $\nu$ is a phenomenological damping coefficient due to local frictional collisions in the dense glassy environment.
To solve this equation, the first step is to take the Fourier transform: $\uline{r}_{i}(t)\rightarrow \tilde{\uline{r}}_i (\omega)$, resulting in the equation:
\begin{align}
-m\omega^2\tilde{\uline{r}}_{i}+i\nu \omega \tilde{\uline{r}}_{i}+\uline{\uline{H}}_{ij}\tilde{\uline{r}}_{j}&=q\widetilde{\uline{E}}.
\end{align}

We then implement normal-mode decomposition: $\tilde{\uline{r}}_i(\omega)=\hat{\tilde{r}}_p(\omega)\mathbf{\uline{v}}_i^p$, where the hat is used to denote the coefficient
of the projected quantity, and $\mathbf{\uline{v}}_i^p$ denotes an eigenvector of the Hessian matrix. Thus the equation of motion is rewritten as
\begin{equation}
-m\omega^2\hat{\tilde{r}}_{p} +i\omega \tilde{\nu}(\omega)\hat{\tilde{r}}_{p}+m\omega_p^2\hat{\tilde{r}}_p =q_{e}\hat{\widetilde{E}}, \notag
\end{equation}
where $\omega_{p}$ denotes the $p$-th normal mode frequency. The equation is solved by 
\begin{equation}
\hat{\tilde{r}}_{p}(\omega)=-\frac{q_{e}\hat{\widetilde{E}}}{m\omega^2-i\omega \tilde{\nu}(\omega)-m\omega_p^2}. \notag
\end{equation}

Upon multiplying through by the eigenvector $\mathbf{\uline{v}}_i^p$, we go back to a vector equation for the Fourier-transformed displacement of particle $i$:
\begin{equation}
\delta\tilde{\uline{r}}_{i}(\omega) =-\frac{q_{e}}{m\omega^2-i\omega\tilde{\nu}(\omega)-m\omega_{p}^2}\widetilde{\uline{E}}(\omega).
\end{equation}

Each particle contributes to the polarization a moment $\uline{p}_{i}=q_{e}\delta\uline{r}_{i}$. In order to evaluate the macroscopic polarization, we need to add together the contributions from all microscopic degrees of freedom in the system, $\uline{P}=\sum_{i}\uline{p}_{i}$.
In order to do this analytically, we use the standard procedure of replacing the discrete sum over the total $3N$ degrees of freedom of the solid with the continuous integral over the eigenfrequencies $\omega_{p}$, $\sum_{p}^{3N}...=\sum_{\alpha=1}^{3}\sum_{i=1}^{N}...\rightarrow \int\rho(\omega_{p})...d\omega_{p}$, which gives the following sum rule in integral form for the polarization in glasses
\begin{equation}
\widetilde{\uline{P}}(\omega)=-\left[\int_0^{\omega_D} \frac{\rho(\omega_p)q_{e}^2}{m\omega^2-i\omega\tilde{\nu}(\omega)-m\omega_p^2}d\omega_{p}\right]\widetilde{\uline{E}}(\omega).
\end{equation}

Here, $\rho(\omega_{p})$ is the vibrational DOS, and $\omega_{D}$ is the cut-off Debye frequency arising from the normalization of the density of states.
The complex dielectric permittivity $\epsilon^{*}$ is defined as $\epsilon^*=1+4\pi\chi_{e}$ where $\chi_{e}$ is the dielectric susceptibility which connects polarization and electric field as~\cite{Born-Wolf}: $\uline{P}=\chi_{e} \uline{E}$.  Hence, we obtain the complex dielectric function expressed as
a frequency integral as
\begin{equation}
\epsilon^*(\omega)=1-\int_{0}^{\omega_{D}}\frac{A\rho(\omega_{p})}{\omega^2-i( \tilde{\nu}(\omega)/m)\omega-C^2\omega_p^2}
d\omega_{p}
\end{equation}
where $A$ is an arbitrary positive constant, $C=\sqrt{\kappa/m}$, and $\omega_{D}$ is the Debye cut-off frequency (i.e.
the highest eigenfrequency in the vibrational DOS spectrum). As one can easily verify, if $\rho(\omega_{p})$
were given by a Dirac delta, one would recover the standard simple-exponential
(Debye) relaxation~\cite{Born-Wolf}.

It is important to emphasize that, in Eq.(5), low-frequency soft modes which
are present in $\rho(\omega_{p})$ necessarily play an important role also at
low applied-field frequencies $\omega$, because of the $\omega^{2}$ term in the
denominator. As we will see below, this fact in our theory implies a direct
role of the boson peak on the $\alpha$-relaxation process (i.e. on the so-called $\alpha$-peak which is the main resonance peak of the loss dielectric modulus).

Since we are using a DOS obtained numerically from a system with a finite
($\sim 4000$) number of particles in our numerical simulation of the model system, it is important to correctly
take care of finite size effects in Eq.(5).
In numerical simulations of a finite system, the DOS
$\rho(\omega_{p})$ is not a continuous function, but discrete. Thus, we rewrite
Eq.(5) as a sum rule over a discrete distribution of $\omega_p$:
\begin{equation}
\epsilon^*(\omega)=1-\sum_{p}\frac{A}{\omega^2-i(\nu/m)\omega-C^2\omega_{p}^2}
\end{equation}
where $A$ has absorbed the scaling constant and we have replaced the density of
states by the discrete sum of $\delta-$functions. Since the dielectric function
is a complex quantity, we can split it into its real and imaginary parts, i.e.
$\epsilon^*(\omega)=\epsilon'(\omega)-i\epsilon''(\omega)$:
\begin{align}
\epsilon'(\omega)&=\epsilon'(\infty)+\sum_p\frac{A_1(C^2\omega_{p}^2-\omega^2)}{(C^2\omega_{p}^2-\omega^2)^2+(\omega\nu/m)^2},
\\
\epsilon''(\omega)&=\sum_{p}\frac{A_2(\omega\nu/m)}{(\omega^2-C^2\omega_{p}^2)^2+(\omega\nu/m)^2},
\end{align}
Here, $A_1,A_2,\epsilon'(\infty)$ are re-scaling constants that have to be
calibrated to adjust the height of the curves.
\\

As remarked above and as observed in numerical calculations of the DOS in the vicinity of the mechanical stability point of disordered solids, there exits
a lowest non-zero eigenfrequency $\omega_{p,min}$ and a vanishingly small gap between $\omega_{p}=0$ and $\omega_{p,min}=0.019$ for $Z=6.1$. Therefore, when $\omega \ll
\omega_{p,min}$,
$\epsilon''(\omega)$ becomes
\begin{equation}
\epsilon''\approx\sum_p\frac{A_{2}\omega\nu/m}{C^4\omega_{p}^4}\sim\omega.
\end{equation}

\section{Results and discussion}
We now present our theoretical predictions based on the Goldhaber-Teller model evaluated with the DOS of Fig.2, and compare them with the response calculated based on the best-matching Kohlrausch stretched-exponential function. In
Fig. 4-5 using Eq. (7)-(8). we plotted the comparisons for $\epsilon'(\omega)$ at $Z=6.1$, ($T=184~K$, i.e.
slightly below $T_{g}$, obtained by implementing the numerical DOS of Fig.2) $Z=7, Z=8$ and $Z=9$ in Eq.(7).
The scaling parameters for $Z=6.1$ are chosen such that $\epsilon'(\omega)$ calibrates with experimental data for glycerol at the glass transition from Ref.~\cite{Loidl} (presented elsewhere). The parameters that control the height of curve (the $A$ parameters) for other $Z$ values are selected such that the curves height is the same. For the other parameter in the model like $C, \nu$, the values are the same for all $Z$.

It can be seen from Fig. 4 that, as $Z$ increases, the curves of $\epsilon'(\omega)$ shift towards the right hand side, with the stretched exponential character of the response decreasing accordingly. Importantly, for $Z=6.1$ the stretched-exponential character of the response is very well recovered by our model over the entire time-domain.
In Fig. 5 we present fittings of the dielectric loss, $\epsilon''(\omega)$.
Also, in this case, the $\alpha$-relaxation wing typical of stretched-exponential relaxation with the paradigmatic exponent $\beta=0.56$ well within the range observed in many systems~\cite{Phillips}, is perfectly reproduced over 6 orders of magnitude in frequency. On the left-hand ascending side of the peak, our model is dominated by the vanishingly small gap between the zero-modes and the lowest boson-peak eigenmode $\omega_{p,min}$, which leads to the power-law,
$\sim\omega^{1}$ as derived in Eq.(9), for the ascending part of the peak.
On the high-$\omega$ side of the peak, where the dynamics is dominated by the
soft boson-peak modes, the DOS is approximately flat as a function of $\omega_{p}$ in Fig.2.

Let us now consider the dielectric loss modulus $\epsilon''(\omega)$.
Upon increasing $Z$, the stretched-exponential character of the response also in this case shrinks in the frequency axis and the response deviates from stretched-exponential at higher frequencies.
This general trend, of stretched-exponential character becoming confined to smaller frequency range upon increasing $Z$, can be explained by recalling that upon increasing $Z$ our model random lattice becomes increasingly more ordered, as reflected in the growing of the Debye-law regime in the DOS and also in the emergence of relics of van Hove singularities. In particular, in this model the increase of $Z$ is accompanied by a marked increase of the local degree of centrosymmetry of nearest-neighbours around a tagged particles as shown in Ref.~\cite{Milkus}.

\begin{figure}
\begin{center}
\includegraphics[height=5.7cm,width=8.5cm]{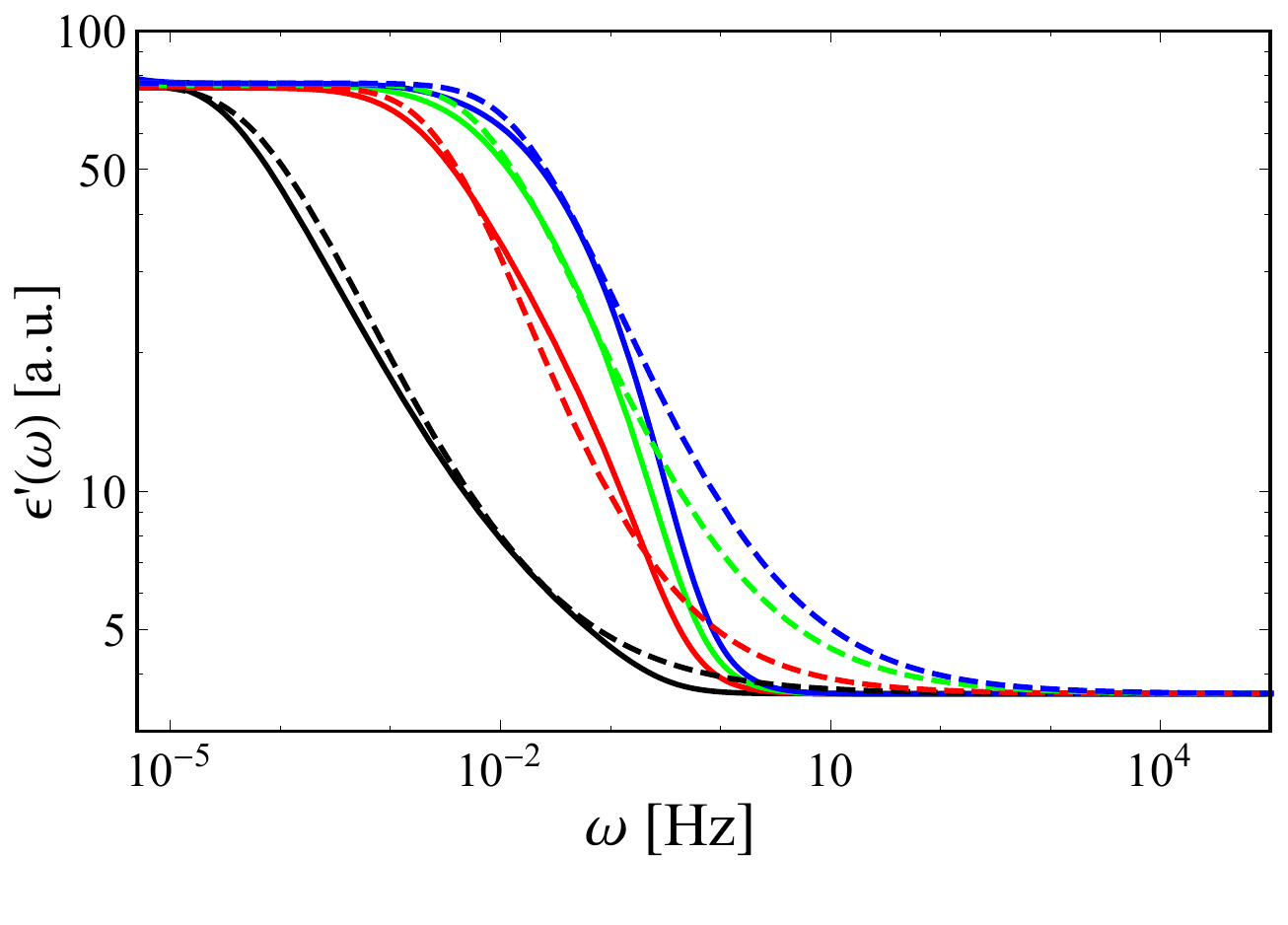}
\caption{Real part of the dielectric function as a function of the frequency $\omega$ of
the applied field. Solid lines are our theory, Eq. (7). The dashed lines are the real part of the Fourier transform when we consider the best-fitting stretched-exponential function with $\beta=0.56, 0.65, 0.60, 0.60, \tau=5655, 155, 50, 25$ respectively. From left to right, each relates to $Z=6.1, 7, 8$ and $9$. We have taken $C=10, \nu/m=1620$. Rescaling constant is used to adjust the height of the curves.}
\end{center}
\end{figure}
\begin{figure}
\begin{center}
\includegraphics[height=5.7cm,width=8.5cm]{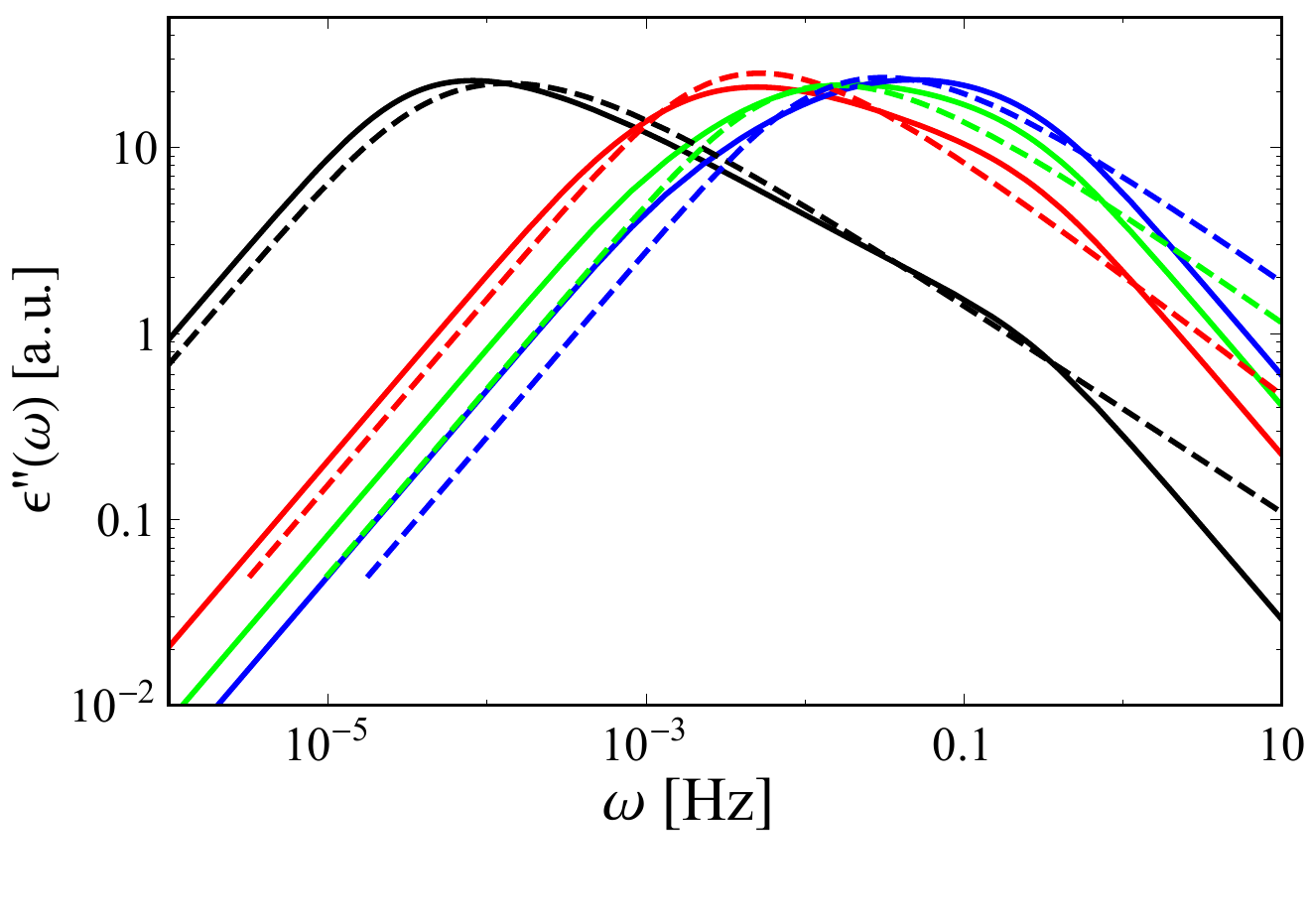}
\caption{Imaginary part of the dielectric function as a function of the frequency of
the applied field. Solid lines are our theory, Eq. (8). The dashed lines are the imaginary part of the Fourier transform when we consider the best-fitting stretched-exponential function with $\beta=0.56, 0.65, 0.60, 0.60, \tau=5655, 155, 50, 25$ respectively. From left to right, each relates to $Z=6.1, 7, 8$ and $9$. We have taken $C=10, \nu/m=1620$. Rescaling constant is used to adjust the height of the curves.}
\end{center}
\end{figure}

Finally, we consider the dielectric response in the time domain. The time dependent
dielectric function $\epsilon(t)$ and complex dielectric function $\epsilon^*(\omega)$ are related
as:
\begin{align}
\frac{d\epsilon(t)}{dt}&=\frac{1}{2\pi}\int_{-\infty}^{\infty}(\epsilon^*(\omega)-\epsilon(\omega=\infty))e^{i\omega t}d\omega \\
\epsilon^*(\omega)&=\epsilon(\omega=\infty)-\int^{\infty}_0\frac{d\epsilon(t)}{dt}e^{-i\omega t}dt.
\end{align}

By using the Fourier inversion theorem, we can write the analytical form of $\epsilon(t)$ as follows:
\begin{multline}
\epsilon(t)=B+{}\\
A\int_0^{\omega_D}
\frac{\rho(\omega_p)}{2K}\left(\frac{e^{(K-\nu/2m)t}}{K-\nu/2m}+\frac{e^{-(K+\nu/2m)t}}{K+\nu/2m}\right) d\omega_p,
\end{multline}
where $K\doteq\sqrt{(C\omega_p)^2-\frac{\nu^2}{4m^2}}$, while $B$ is a re-scaling constant.
This equation is a key result: it provides a direct and quantitative relation
between the macroscopic relaxation function of the material and the DOS. As we show below, the
presence of a boson peak in $\rho(\omega_{p})$ directly causes
stretched-exponential decay in $\epsilon(t)$ via Eq.(12). Further, if the DOS is nothing but a Dirac delta centered at a given Einstein frequency $\omega_E$ (as in the Einstein model of solids~\cite{Born-Huang}), then the integral of Eq. (5) gives, for the imaginary part, a simple Lorentzian curve with a symmetric peak centered on $\omega_E$ which is the resonance frequency of the system. Upon performing the inverse Fourier transform, one recovers a simple exponential relaxation in the time domain, i.e. $\epsilon(t)\sim \exp(-t/\tau)$ with some characteristic time $\tau$, as in Debye relaxation~\cite{Donth}.

In Fig. 6, we plot predictions of Eq.(12) with the parameters calibrated in the
glycerol data fitting, along with the Kohlrausch function~\cite{note,Montroll},
for the relaxation in the time domain.
It is seen that for $Z=6.1$ our theory based on soft modes is able to perfectly recover
stretched-exponential relaxation, with stretching-exponent $\beta=0.56$, over the entire time-domain. Without the
boson-peak modes in the DOS, we have checked that stretched-exponential relaxation cannot
be recovered at all, and the decay is simple-exponential. Hence, our Eq.(12) provides
the long-sought cause-effect relationship between soft modes and
stretched-exponential relaxation.
For higher values of $Z$, also in this case the stretched-exponential character of our theoretical model decreases, although for e.g. $Z=7$ the relaxation predicted by our theory is still very close to stretched-exponential.

\begin{figure}
\begin{center}
\includegraphics[height=5.7cm,width=8.5cm]{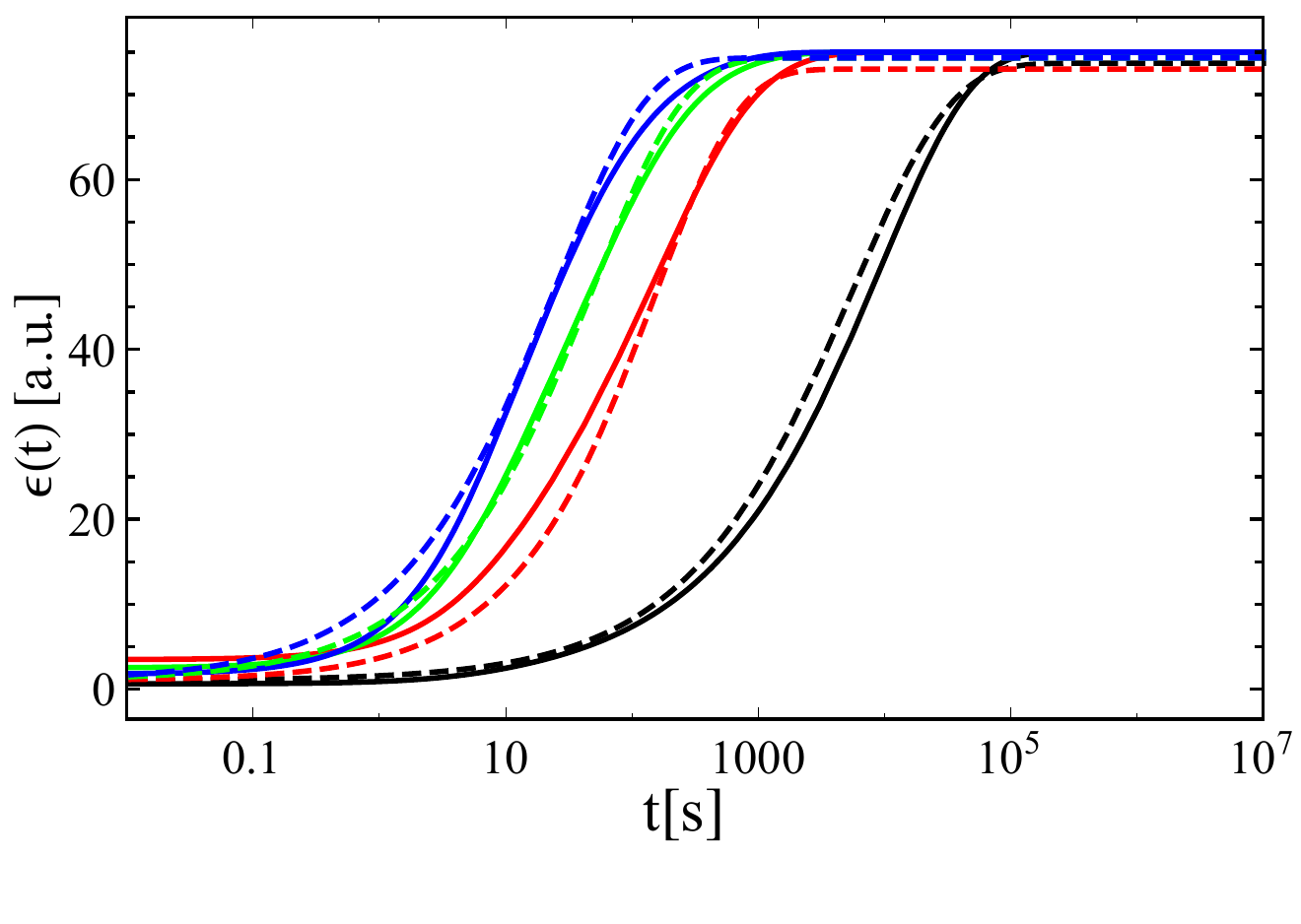}
\caption{Time-dependent dielectric response. Solid lines are calculated from
our theory using Eq.(12) with physical parameters calibrated in the fitting of
Fig.3. The dashed lines represent the stretched-exponential Kohlrausch function with $\beta=0.56, 0.65, 0.60, 0.60, \tau=5655, 155, 50, 25$ respectively, that are closest to our theoretical curves. From right to left side,  $Z=6.1, 7, 8, 9$. Rescaling constants are used to adjust the height of the curves.}
\end{center}
\end{figure}

\section{Conclusion}
We have studied a classical Goldhaber-Teller model of a disordered dielectric where particles
are placed at random in the lattice and interact harmonically with nearest-neighbours of opposite charge by making two assumptions in our model system: (1) Electric charges are elastically bonded by harmonic springs. (2) The damping force with constant damping coefficient is assumed in the particle motion. The assumption (1) of electric charges being bound to other nearby opposite charges via harmonic springs is essential to describe the dynamics via the Hessian matrix and normal modes. This assumption is an approximation for dielectrics, e.g. organic glasses, where the partially negatively charged segments (which bear a partial negative charge) are elastically bound to the partially positively charged segments via attractive interactions. In particular, a partially negatively charged segment is bound via covalent bonding plus electrostatic attraction to a positively charged segment of the same molecule. It can also be attracted to a positively charged segment of a different molecule via either electrostatic attraction or van der Waals attraction, or both. These attractive interactions are typically longer ranged (e.g. screened-Coulomb attraction), whereas repulsive steric interaction due to closed-shell electron Pauli repulsion are short-ranged and typically decay exponentially according to quantum mechanics. Hence a charged segment occupies a local minimum of interaction energy. In the linear response regime, where only small displacements from the minimum are considered, the charge can be approximated as a harmonic oscillator, independently of the precise shape of the interaction energy landscape but provided that the minimum is steep enough. The same approximation may apply to ionic liquids. The assumption (2), which introduces a damping coefficient in the equation of motion, is an assumption of the Lorentz oscillator model in electrodynamics. This assumption accounts for the drag force that a charge encounters while it oscillates in the potential minimum well. This drag force, which effectively dampens the motion, is due to the collisions and to the coupling with other molecular segments or other vibrating molecules nearby. Clearly, this is a phenomenological coefficient, which however can be derived from nonequilibrium statistical mechanics, e.g. from so-called Caldeira-Leggett models.

The vibrational density of states of disordered solids is dominated
by an excess of low-energy modes that contributes the so-called boson-peak above the Debye law in the DOS.
Working with the standard Lorentz sum rules and implementing the DOS of a model disordered solid, we have derived expressions for the dielectric function and
the dielectric relaxation which account for the disordered structure.
In this model, the coordination number $Z$ controls the extent of the boson peak and the vicinity to the Maxwell rigidity transition $Z=6$  (that
could be identified with a glass transition if $Z$ is related to temperature, e.g. via thermal expansion).
For the condition closest to the transition that we studied, we found strikingly that the dielectric response of our model system exhibits perfect
Kohlrausch stretched-exponential behaviour in both frequency and time domain, over the entire range of frequency and time.
The stretched-exponential character decreases with increasing the average number of nearest-neigbours $Z$ and gets confined to smaller range of frequency/time due to the fact that structural order (in particular, the local centrosymmetry~\cite{Milkus}) increases upon increasing $Z$ in our model.
This framework shows, for the first time, that stretched-exponential relaxation in glasses is directly caused by
the quasi-localized boson-peak modes contribution to the lattice dynamics.
These results open up new opportunities to understand the crucial link between
$\alpha$-relaxation, boson peak and dynamical
heterogeneity in glasses.

\begin{acknowledgements}
Many useful discussions with R. Richert, W. Goetze, and with E. M. Terentjev are gratefully acknowledged.
\end{acknowledgements}


\begin{thebibliography}{99}
\bibitem{note}
The response in the time domain is related to the response in the frequency
domain~\cite{Montroll}
\begin{align}
\epsilon^*(\omega)&=\epsilon'(\omega)-i\epsilon''(\omega) \notag \\
&=\epsilon^*(\infty)+(\epsilon^*(\infty)-\epsilon^*(0))\int_{0}^{\infty}
e^{-i\omega t}\frac{d\phi}{dt} dt.
\end{align}
The relaxation function is related to the response in the time domain
via~\cite{Montroll}:
\begin{equation}
\begin{split}
\epsilon(t)&=\epsilon^*(\omega=\infty)+{}\\
&(\epsilon^*(\omega=0)-\epsilon^*(\omega=\infty))(1-\phi(t))
\end{split}
\end{equation}
where $\phi(t)=\exp{[-(t/\tau)^{\beta}]}$.
Predictions of this equation have also been plotted for comparison in Fig. 2-3, for the frequency-dependent response, and in Fig. 4 for the time-dependent response, together with the predictions of our theory.
Note that the parameters in the Kohlrausch model have been also adjusted to
match with the experimental data.
\end{thebibliography}
\end{document}